\documentclass[10pt,letterpaper]{article}
\usepackage{opex3}
\usepackage{epstopdf}
\usepackage{mathptmx}
\usepackage{cite}

\begin{document}

\title{Single-mode optical fiber for high-power, low-loss UV transmission}

\author{Yves Colombe*, Daniel H. Slichter*,  Andrew C. Wilson, Dietrich Leibfried and David J. Wineland}

\address{ National Institute of Standards and Technology, 325 Broadway, Boulder, CO 80305 USA}
\address{* See author contribution statement in acknowledgments}

\email{dhs@nist.gov} 

\begin{abstract*}
We report large-mode-area solid-core photonic crystal fibers made from fused silica that resist ultraviolet (UV) solarization even at relatively high optical powers.  Using a process of hydrogen loading and UV irradiation of the fibers, we demonstrate stable single-mode transmission over hundreds of hours for fiber output powers of 10 mW at 280 nm and 125 mW at 313 nm (limited only by the available laser power).  Fiber attenuation ranges from 0.9 dB/m to 0.13 dB/m at these wavelengths, and is unaffected by bending for radii above 50 mm.   
\end{abstract*}



\section{Introduction}
Single-mode optical fiber is an essential and widely-used tool at visible and infrared (IR) wavelengths, but is not generally available for applications in the ultraviolet (UV).  A major cause of this deficit is the phenomenon of UV-induced color-center formation \cite{Skuja2001} \textemdash known as UV solarization\textemdash which causes substantial attenuation of the transmitted light, often to the point where the fiber becomes unusable.  Depending on wavelength, solarization can occur even at microwatt power levels in fused silica fibers.

UV light can create color centers by interacting with dopants or impurities present in the silica glass \cite{Skuja2001}, as well as with strained molecular bonds created during the fiber drawing process \cite{Friebele1976, Hanafusa1987}. Since germanium-doped silica is especially prone to UV solarization, commercially available multimode step index UV fibers typically have an undoped silica core and fluorine-doped cladding \cite{Klein2013}.  Silica with a high hydroxide (OH) content is often used for the core, as it provides some improved resistance to solarization \cite{Skuja2001, Kuzuu1995}. The presence of dissolved molecular hydrogen has also been shown to mitigate the formation of color centers in silica \cite{Faile1967, Shelby1979}.  Hydrogen loading of fibers by prolonged exposure to high-pressure hydrogen gas can be combined with exposure to UV light to confer high solarization resistance \cite{Karlitschek1998, Karlitschek1998a, Oto2001}.  The UV exposure step converts defect precursors in the fused silica into color centers, which then react with the hydrogen and are converted to new forms that are either optically inactive or interact with different wavelengths than the original color centers \cite{Ikuta2002}.  For wavelengths between roughly 200 nm and 400 nm, this process results in reduced fiber attenuation.  A similar process of hydrogen loading and gamma ray exposure has been used to ``harden'' fibers against degradation caused by ionizing radiation \cite{Lyons1992,Tomashuk1998,Henschel2002, Griscom2013}.

To date, these methods have been successfully employed with multi-mode fibers, but single-mode step-index fibers for the UV present additional challenges.  The small mode field diameter of such fibers (typically $\sim 2\; \mu$m) results in high UV intensity in the core, giving rise to solarization at lower transmitted powers.  Furthermore, in-coupling to such small cores requires the use of complex optics to achieve good mode matching.  One solution to these issues is to use photonic crystal fiber (PCF) with a large mode area (LMA).  The large mode area decreases the intensity in the core for a given transmitted power, and also allows the use of a single spherical lens for efficient in-coupling.  Such fibers can exhibit ``endless'' single-mode operation \cite{Birks1997} extending from the IR to the UV, which is beneficial for applications requiring co-propagating laser beams of disparate wavelengths.  

Several studies have been performed on large-mode-area PCF with UV and near-UV light.  In 2009, Yamamoto \textit{et al.} \cite{Yamamoto2009} investigated the performance of LMA-10-UV fiber (NKT Photonics \cite{footnote1}) at 250 nm.  With $\sim0.3$ mW incident upon a 1.3 m length of fiber, they observed a steady drop in the transmission during the first 20 hours to 65\% of its initial value, followed by stable transmission.  For higher optical powers, stable transmission was not observed, and at 3 mW input power, the transmission was reduced to a few percent in less than 5 hours. In 2010, Gonschior \textit{et al.} \cite{Gonschior2010} demonstrated stable transmission of 405 nm light through LMA-10-UV fiber at output powers of up to 105 mW over a period of weeks.  Fused silica is generally very resistant to solarization at this longer wavelength \cite{Skuja2001}.  

In this paper, we report the investigation of hydrogen loading and UV irradiation of photonic crystal fibers as a method for producing solarization-resistant single-mode fiber for the UV.  While such fibers may have broad applicability, our group is primarily interested in their use in laser spectroscopy of trapped ions.  In particular, quantum information experiments with trapped ions \cite{Leibfried2003,Monroe2013} require high power beams in the UV\textemdash in our case, beams with up to tens of mW of 280 nm and 313 nm light\textemdash with clean mode profiles and minimal beam-pointing fluctuations.  For this task, solarization-resistant single-mode UV fibers would be a natural solution.  

\section{Fiber Preparation and Experimental Setup}

\begin{figure}[tbp]
\centering\includegraphics{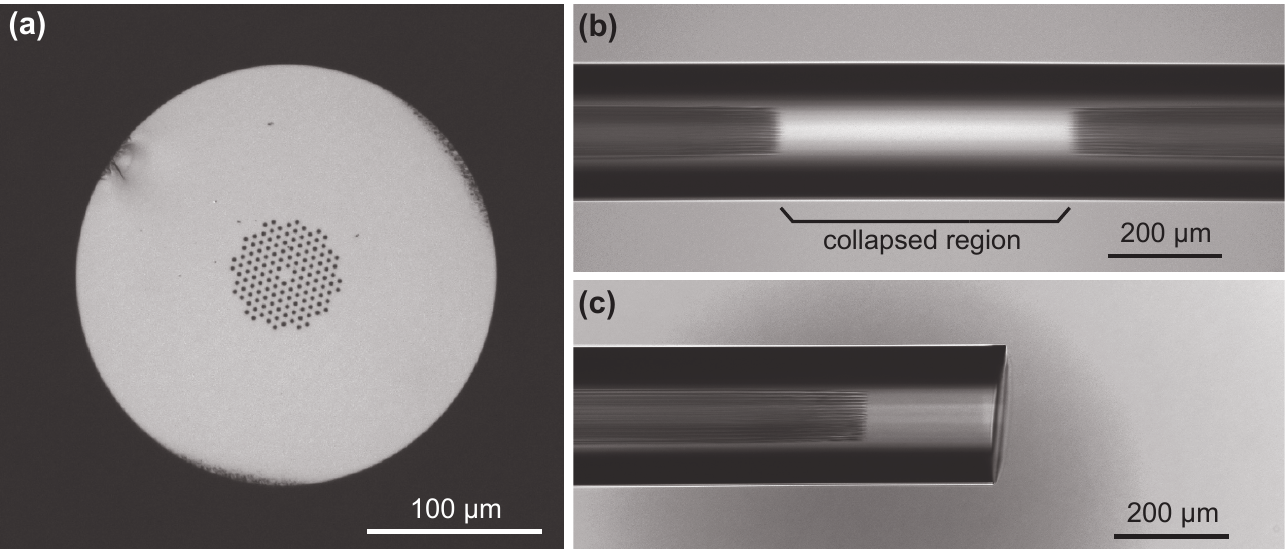}
\caption{\label{fiber} Solid-core photonic-crystal fibers.  Panel (a) shows an optical micrograph of the cleaved facet of an uncollapsed LMA-8-UV fiber.  The pattern of holes which forms the photonic crystal can be seen in the center.  The guided mode propagates through the solid core in the center of the array of holes.  The cross-section of LMA-10-UV fiber looks qualitatively the same.  Panel (b) shows a side view of a fiber where the photonic crystal has been collapsed over a 600 $\mu$m section in the center using a fusion splicer.  The collapsed region appears bright, with the dark photonic crystal holes visible on either side.  In (c), a fiber is shown after angle cleaving in a collapsed section.}
\end{figure}

We investigated two types of solid-core PCF, LMA-10-UV and LMA-8-UV (NKT Photonics).  Both fibers have a hexagonal lattice of air holes surrounding a solid core.  LMA-10-UV is a stock fiber made from low-OH fused silica (Heraeus Quarzglas F300). It has a core size of $10 \pm 1\; \mu$m and a cladding diameter of $230\pm 2\; \mu$m.  LMA-8-UV is a custom fiber made from high-OH fused silica (Heraeus Quarzglas F110), with a core size of $8.6 \pm 0.5\; \mu$m and a cladding diameter of $240 \pm 2\; \mu$m.  The cladding diameter of these fibers is larger than that of typical single-mode fibers; this reduces micro-bending, which can increase propagation losses at short wavelengths \cite{Nielsen2003}.   An optical micrograph of the fiber cross-section for LMA-8-UV is shown in Fig. \ref{fiber}(a).

The fibers were hydrogen-loaded by exposure to hydrogen gas (99.95\%) at room temperature and a pressure of $\sim$ 10 MPa for a duration of 4-6 days.  Hydrogen loading was performed by O/E Land, Inc., and also on-site at the NIST Hydrogen Fuel Materials Test Facility.  To reduce out-diffusion of hydrogen after loading, the loaded fibers from O/E Land were shipped in dry ice and stored in a freezer at $-85^{\circ}$C until use.  At this temperature, the diffusion coefficient of H$_2$ in fused silica is calculated to be a factor of 25,000 times smaller than at room temperature \cite{Shang2009}.  Fibers loaded at NIST were placed in storage at $-85^{\circ}$C within two hours of removal from the hydrogen chamber.  

Hydrogen-loaded fibers were allowed to warm to room temperature and then connectorized at one or both ends to form patch cables for testing.  To connectorize a fiber, we first collapsed the photonic-crystal holes in a 400- to 700-$\mu$m-long section near the end of the fiber using a fusion splicer arc, as shown in Fig. \ref{fiber}(b).  Because the holes represent a relatively small fraction of the total cross-sectional area, we found that the outer diameter of the fiber changed negligibly upon collapsing the holes inside.  We then angle cleaved the fiber ($\approx5^{\circ}$ facet angle) in the collapsed region, leaving a collapsed section of approximately 200-300 $\mu$m at the end of the fiber, seen in Fig. \ref{fiber}(c).  We used a cleaver designed to produce controlled angle cleaves on large-diameter fibers (Fujikura CT-100).  The angle cleave increases the return loss at the fiber output by at least 40 dB relative to a straight cleave.  We found that fibers without a sufficiently angled cleave at the output showed some back-reflection and tended to form narrowband reflective gratings after several days of UV exposure.  This effect was more pronounced for LMA-8-UV fibers than for LMA-10-UV fibers.  After cleaving, the end of the fiber was mounted in an FC/PC connector with a zirconia ferrule, supported and protected by stainless steel and polyimide tubing.  The assembly was held together with low-stress UV-curable adhesive (Norland NOA 65) applied approximately 40-50 mm from the fiber facet to minimize any detrimental effects on the mode quality and transmission due to fiber stress from gluing \cite{footnote2}.  

For some measurements (notably the cutback measurements) we cleaved the fiber output without collapsing the photonic crystal.  However, we generally collapsed the fibers before cleaving when they would be connectorized, as it is easier to clean the facet of a collapsed fiber than an uncollapsed one (where debris on the facet can collect in the photonic crystal holes).  Collapsing the ends is also expected to slow the out-diffusion of hydrogen, and can help reduce UV-induced carbon deposition on the input facet, since the input light is not as strongly focused there \cite{Gonschior2010}.  The presence or absence of a collapsed region at the input facet did not appear to affect fiber coupling efficiency.  The output mode quality did not depend strongly on whether or not the fiber was collapsed at the output, but uncollapsed fibers tended to show a larger number of low-intensity features surrounding the main output peak.  

Once connectorized, the hydrogen-loaded fibers were placed into the optical setup at room temperature and UV light was coupled in to ``cure'' the fiber.  Because the hydrogen diffuses out of the fiber more rapidly at room temperature, we began the curing process as soon as possible after removing the fibers from cold storage.  In our experiments, all fibers began curing within three hours (and most within one hour) of removal from the freezer.  This interval is much shorter than the time scale for hydrogen out-diffusion at room temperature, which has been estimated to be between several days and several weeks \cite{Lyons1992, Henschel2002}.  A simple calculation using the measured hydrogen diffusion rate from the literature \cite{Shang2009} predicts that the hydrogen concentration in the center of the fiber will be halved from its initial value after approximately one month at room temperature assuming a solid fused silica fiber (approximating a fiber collapsed at both ends), or after about two to three hours at room temperature with a zero-hydrogen boundary condition at the radius of the innermost air holes (approximating an uncollapsed fiber).  

We tested the fibers using both 313 nm and 280 nm laser light.  The 313 nm light was generated by first frequency summing two infrared fiber lasers (at 1051 nm and 1550 nm) to generate 626 nm light and then frequency doubling this light with a resonant nonlinear cavity, as described in \cite{Wilson2011c}. The maximum operating power was approximately 200 mW.  The 280 nm light was derived from a fiber laser at 1118 nm followed by two consecutive stages of frequency doubling, and had a maximum operating power of approximately 15 mW. The output modes of both sources are slightly elliptical, with typical beam diameters at $1/e^2$ intensity of 450-600 $\mu$m.  The UV light was focused into the fibers using an anti-reflection-coated fused silica biconvex lens with $f=20$ mm; no spatial filtering or cleaning of the doubler output mode was performed prior to the fiber-coupling stage.  

Input power was monitored using a 5\% pickoff (or a 50-50 beamsplitter for the cutback measurements) and a UV-extended Si photodiode power sensor.  The output power was measured using either a UV-extended Si photodiode power sensor or a thermopile power sensor, depending on the output power level.  The exact pickoff ratio and mutual detector linearity were calibrated for each test fiber by placing the output power sensor in the input beam just after the fiber in-coupling lens and recording readings from both input and output sensors over a range of input powers.  

\section{\label{perftest} Results}

\subsection{Mode shape and higher-order modes}

\begin{figure}[htbp]
\centering\includegraphics{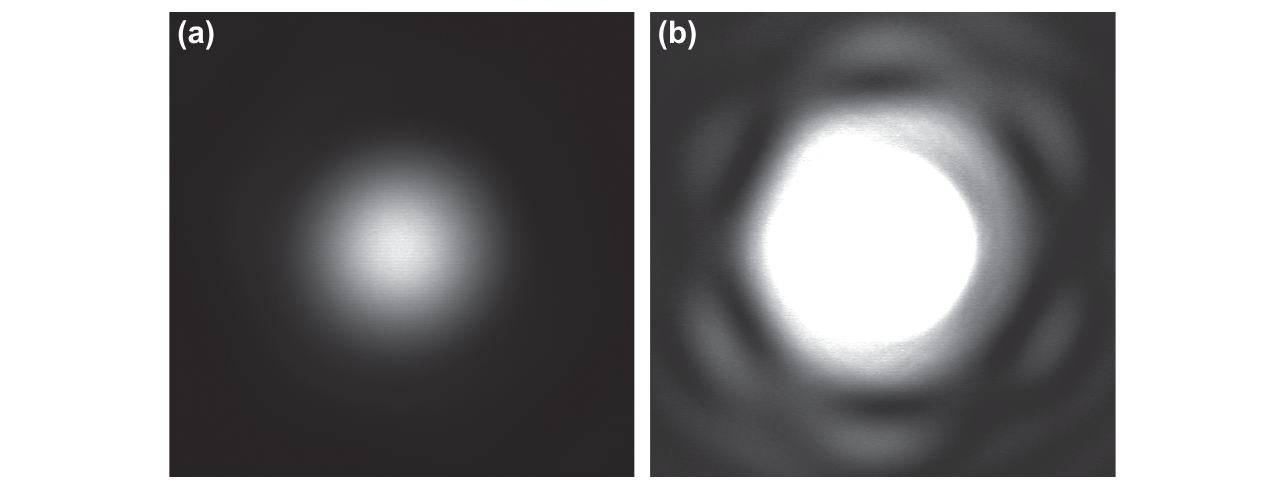}
\caption{\label{mode} Far-field output-mode spatial intensity distribution at 313 nm. Panel (a) shows a typical output beam shape for a collapsed, cured fiber (here, an LMA-10-UV fiber).  Panel (b) shows the same beam with increased detector gain.  The central peak has saturated the detector, and weak hexagonally-arranged sidelobes are visible around the main peak.}
\end{figure} 

The far-field spatial output mode of a collapsed, cured LMA-10-UV fiber at 313 nm is shown in Fig. \ref{mode}.  The mode has a $95\,\pm1$\% overlap with a Gaussian profile, calculated using a power-in-the-bucket method \cite{Siegman1998}. The main deviation from a Gaussian shape comes from low-intensity sidelobes in a hexagonally symmetric pattern, which contain roughly 5\% of the output power. These sidelobes can be seen in Fig. \ref{mode}(b), where the gain on the beam profiling camera has been increased relative to Fig. \ref{mode}(a) to highlight the sidelobes.  If desired, the sidelobes can be removed using an appropriate circular aperture without substantially clipping the main peak.   At 313 nm, the measured beam divergence half-angle of the output mode at $1/e^2$ intensity was $1.16^\circ\pm0.02^\circ$ for LMA-10-UV and $1.45^\circ\pm0.05^\circ$ for LMA-8-UV.  

Both types of fiber are also capable of transmitting higher-order modes in the UV, most noticeably very high order ``speckle'' modes which appear to propagate through the cladding or holey region.  However, when the input is well-coupled to the lowest-order mode, these higher modes are not excited, even with intentional lateral misalignment of the input beam.  As an additional precaution, bends can be introduced into the fiber to preferentially extinguish transmission of the higher-order modes.  Gentle bends ($\sim$ 1 m radius) are sufficient to suppress the propagation of most higher-order modes.  In general, a single $180^\circ$ bend of radius 6-10 cm somewhere along the fiber ensures that only the desired lowest-order mode will propagate.  

For unexplained reasons, the presence of hydrogen in the fiber appears to increase the transmission and/or in-coupling of ``speckle''-type modes in the UV.  Finding and coupling into the lowest-order mode of a hydrogen-loaded fiber to begin the curing process can be challenging, but once the fiber has cured for several days this phenomenon is no longer evident.  

\subsection{Fiber attenuation}

\begin{figure}[htbp]
\centering\includegraphics{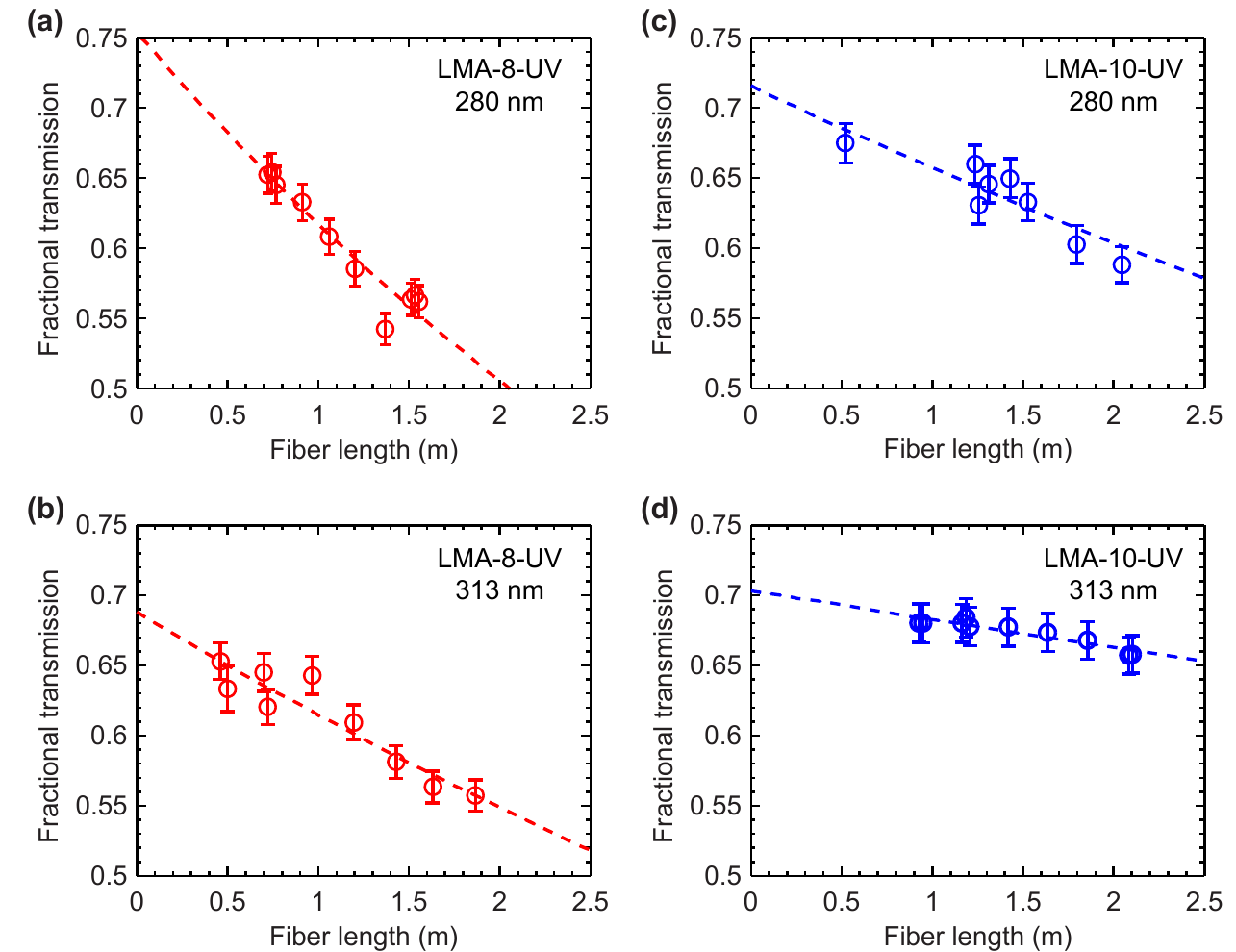}
\caption{\label{cutback} Cutback measurements of cured fibers. The four panels show cutback data for LMA-8-UV at 280 nm (a) and 313 nm (b), as well as for LMA-10-UV at 280 nm (c) and 313 nm (d).  Error bars are dominated by uncertainties from the power sensors.  The dotted lines are fits to a decaying exponential, yielding attenuations per unit length of $0.9\pm0.2$ dB/m (a), $0.5\pm0.1$ dB/m (b), $0.4\pm0.2$ dB/m (c), and $0.13\pm0.04$ dB/m (d).  Extrapolation of the fits to zero length gives typical estimated in-coupling efficiencies around 0.7.}
\end{figure}

Fiber transmission losses were measured by a standard cutback technique.  The hydrogen-loaded fibers were connectorized at the input end and cured with UV light at 1.5 to 2 mW output power for 16 to 20 hours.  A bend of roughly 60 mm radius near the input suppressed the transmission of any modes other than the lowest-order mode.  The output end of the fiber was not collapsed, and the fiber was progressively shortened with a series of {\it in situ} cleaves from an initial length of between 1.5 and 2.2 m.  Measurements were performed with typical output powers of 1 to 1.5 mW.  The transmission of the fiber was measured at each length, with the data shown in Fig. \ref{cutback} along with least-squares fits to a decaying exponential.  The \textit{y}-intercept of the fit gives the in-coupling efficiency (including reflection losses from both facets), which was typically around 0.7.  

For LMA-8-UV, the loss is $0.9\pm0.2$ dB/m at 280 nm and $0.5\pm0.1$ dB/m at 313 nm.  For LMA-10-UV, the loss is lower, $0.4\pm0.2$ dB/m at 280 nm and $0.13\pm0.04$ dB/m at 313 nm.  

\begin{figure}[htbp]
\centering\includegraphics{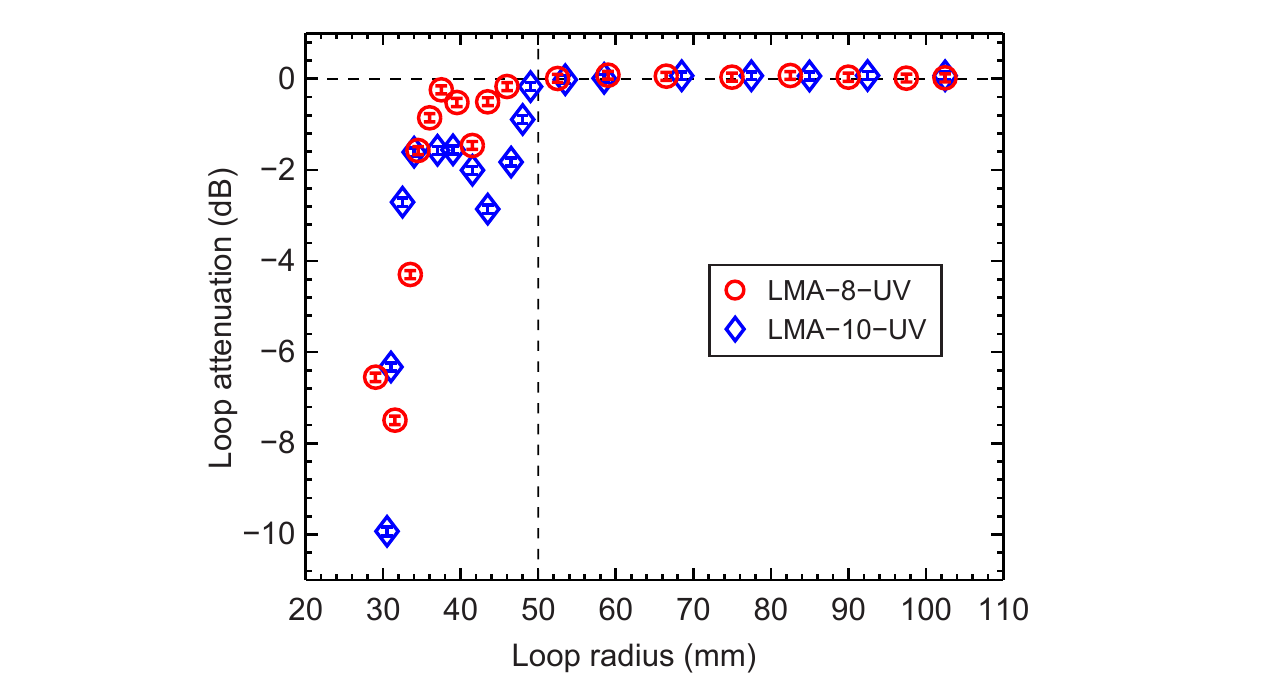}
\caption{\label{bend} Bending loss.  Transmission loss at 313 nm due to the inclusion of a single loop of constant radius in LMA-8-UV (red circles) and LMA-10-UV (blue diamonds) fibers.  Dashed lines indicating 0 dB loss and 50 mm loop radius are shown as guides to the eye.  Both fibers show a revival of transmission near 38 mm loop radius.  The transmitted mode is observed to be the lowest-order mode at all radii.}
\end{figure}

We measured bending losses in the fibers by examining transmission at 313 nm as a function of the radius of a single loop introduced into the fiber.  Data were taken for decreasing loop radii until transmission dropped to zero, occurring at $27\pm 1$ mm radius for LMA-8-UV and $29.5\pm 1$ mm radius for LMA-10-UV.  The results are shown in Fig. \ref{bend}.  Both LMA-8-UV and LMA-10-UV show negligible bending loss for bend radii above 50 mm, simplifying the task of routing fibers used for laboratory applications.  

\subsection{High-power tests}

\begin{figure}[tbp]
\centering\includegraphics{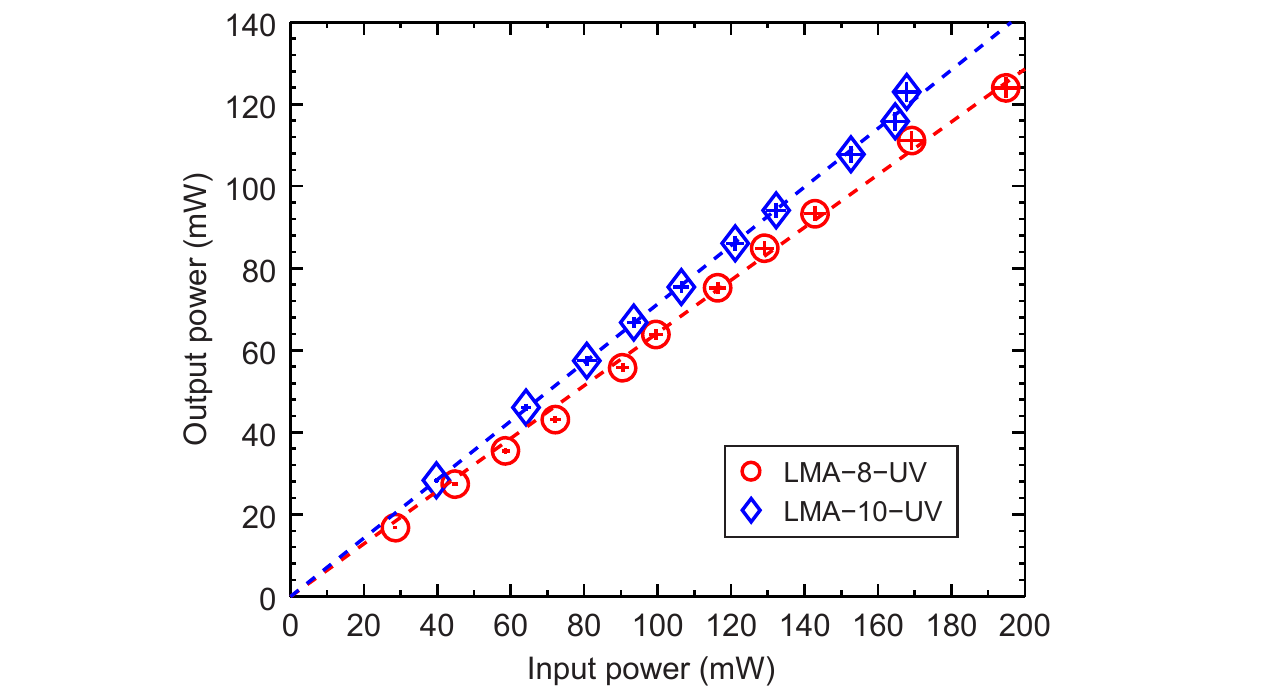}
\caption{\label{powlin} Transmission linearity.  Output power is plotted versus input power for $\sim$ 1 m lengths of LMA-8-UV (red circles) and LMA-10-UV (blue diamonds) fibers. Both fibers were cured at $\sim$ 100 mW output power.  Dashed lines are linear fits through the origin, corresponding to fractional transmission of 0.64 and 0.71, respectively (including all coupling losses).  Measurements were made at 313 nm.}
\end{figure}

We first characterized the high-power performance of the fibers by examining the power dependence of the transmission.  For uncured, hydrogen-loaded fibers roughly 1 m in length, the output power depends linearly on the input power up to around 100 mW output at 313 nm, where it saturates.  The saturation of the output power was not accompanied by increased back-reflection, as would arise from stimulated Brillouin scattering.  Following several days of curing at $\sim$ 100 mW output power, this saturation effect is no longer observed and the transmission becomes independent of input power up to the highest powers measured, as shown in Fig. \ref{powlin}.  The curing process also appears to improve transmission by a few percent over the first 30-50 hours.  We note that for the high-power tests we cured the fiber at considerably higher powers than was done for the cutback measurements.  

In Fig. \ref{powlin} we plot output power as a function of input power (using 313 nm light) following curing for $\sim$ 1 m lengths of both types of fiber, showing a linear relationship up to the maximum available input power.  We note that this maximum power is different for the two fibers due to drifts in the 313 nm source.  The slope of the fit line gives the fractional fiber transmission, which is 0.64 for the LMA-8-UV fiber and 0.71 for the LMA-10-UV fiber, including all coupling losses.

\begin{figure}[tbp]
\centering\includegraphics{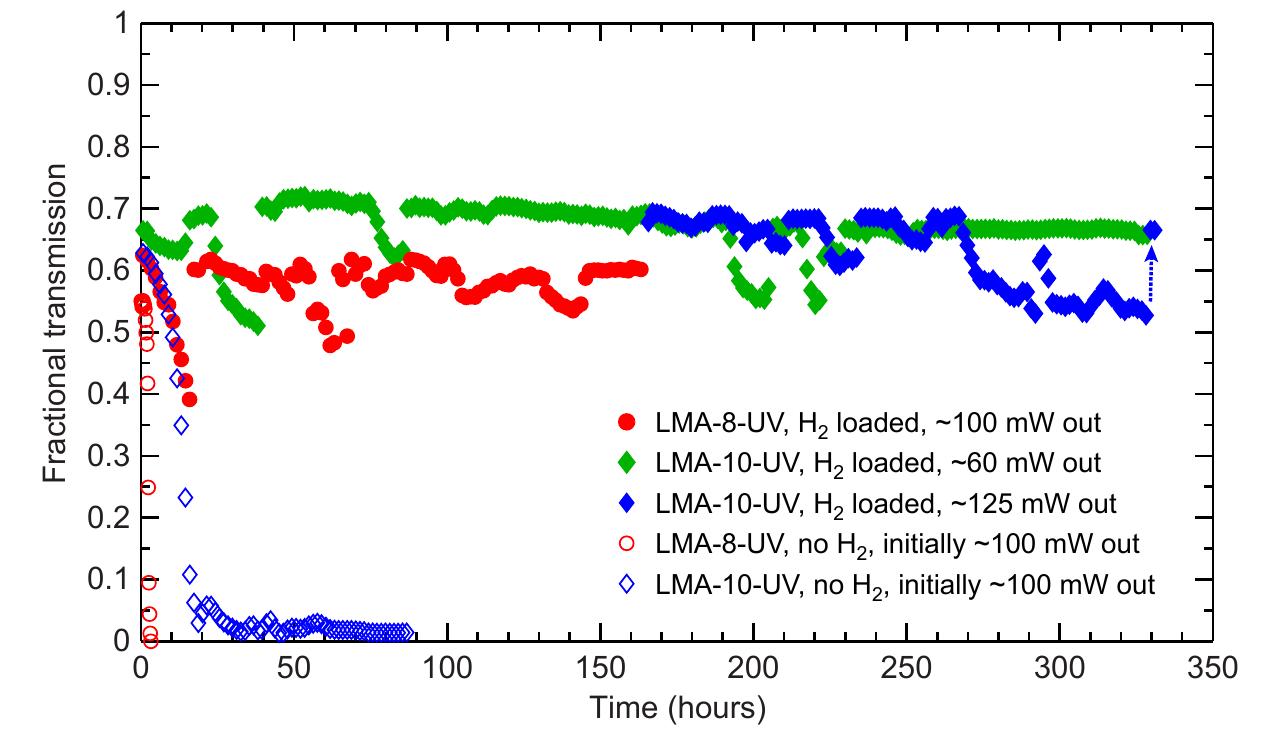}
\caption{\label{highpow} Long-term transmission with high-power 313 nm light.  Transmission is plotted for three different hydrogen-loaded fibers: LMA-8-UV fiber exposed to $\sim$ 100 mW output power continuously (red solid circles), LMA-10-UV fiber cured for 40 hours at $\sim$ 100 mW output power and then exposed to $\sim$ 60 mW output power (green solid diamonds), and LMA-10-UV fiber cured for 40 hours at $\sim$ 100 mW output power and then exposed to $\sim$ 125 mW output power (blue solid diamonds). Transmission is also plotted for two non-hydrogen-loaded fibers exposed with an initial output power of $\sim$ 100 mW: LMA-8-UV (red open circles) and LMA-10-UV (blue open diamonds).  The time axis is referenced to the start of the exposure, including curing.  Data for the first 160 hours of the highest power LMA-10-UV (blue solid diamonds) are not shown.  Drifts in the transmission are due to drifts in the input coupling.  Sudden upward jumps in the measured transmission are due to re-optimization of the fiber in-coupling.  The arrow at far right indicates where the input coupling for the last fiber (blue solid diamonds) was re-optimized shortly before the end of the test.}
\end{figure}

We also monitored the transmission of $\sim$ 1 m lengths of hydrogen-loaded fiber over time during continuous UV exposure to check for degradation from cumulative dose effects or hydrogen out-diffusion.  Figure \ref{highpow} shows data for three hydrogen-loaded fibers at 313 nm: an LMA-8-UV fiber exposed to $\sim$ 100 mW output power continuously for one week (including curing time), an LMA-10-UV fiber cured at $\sim$ 100 mW output power for 40 hours and then exposed to $\sim$ 60 mW output power for 12 more days, and an LMA-10-UV fiber cured at $\sim$ 100 mW output power for 40 hours and then exposed to $\sim$ 125 mW output power for 12 more days.  All three fibers were collapsed at both ends, and exhibited fairly constant fractional transmission of between 0.6 and 0.75 for the duration of the tests.  The second fiber was stored for 38 days at room temperature following this measurement, after which its high-power transmission was again measured with $\sim$ 60 mW output power over 90 hours and found to be unchanged, a total of 8 weeks after the start of the curing process.  In a different test, we baked a collapsed, 280-nm-cured LMA-8-UV fiber at $140^\circ$C for 113 hours (equivalent to 2.6 years of out-diffusion at room temperature) to remove any residual hydrogen and found no solarization after 14 hours of subsequent exposure to 280 nm light with $\sim$ 2 mW output power.

Non-hydrogen-loaded fibers of $\sim$ 1 m length were also tested with $\sim$ 100 mW initial output power at 313 nm to confirm the importance of hydrogen for the curing process, as seen in Fig. \ref{highpow}.  A non-loaded LMA-8-UV fiber solarized to zero transmission in 4 hours under these conditions.  A non-loaded LMA-10-UV fiber with the same initial output power decayed to fractional transmission of around 0.05 over the course of 20 hours and leveled off at a fractional transmission of $\sim$ 0.015 after 60 hours.  We also baked an uncured, hydrogen-loaded, non-collapsed LMA-8-UV fiber for 133 hours at $140^\circ$C to remove the hydrogen prior to UV exposure.  When exposed to 280 nm light with $\sim$ 200 $\mu$W initial output power, this fiber solarized as rapidly as a non-loaded fiber, suggesting that hydrogen loading does not confer resistance to solarization unless UV curing occurs before the hydrogen has diffused out.  

We also performed long-term high-power tests on LMA-8-UV fiber at 280 nm.  We cured the fiber for 95 hours at $\sim$ 2.5 mW output and then increased the output to $\sim$ 10 mW (the maximum achievable with our 280 nm laser system) for another 100 hours.  The transmission was comparable to that seen in the 313 nm tests and remained stable throughout this entire period.  Non-loaded LMA-8-UV fiber was also tested with 300 $\mu$W of input power at 280 nm, solarizing to fractional transmission of $\sim$ 0.2 in one hour and remaining at that level.  

Figure \ref{highpow} also shows some regions where the fiber transmission fluctuates substantially on 5-10 hour timescales.  This effect arises from drifts in the input coupling, which is very sensitive to misalignment due to the small acceptance angle of the fiber.  In all instances, the transmission can be recovered (abrupt upward jumps) by simply re-optimizing the in-coupling to the fiber.  Drifts were seen primarily in the lateral alignment of the input beam on the fiber facet, which were corrected with minute adjustments of the final mirror.  Similar in-coupling drifts have been seen by others working with LMA-10-UV fiber \cite{Gonschior2010}.  We also observed shifts in the optimal lens-fiber distance, which increased by roughly 200 $\mu$m over the first one to two days of curing and then stabilized. This shift occurred gradually over tens of hours and remained even if the input light was temporarily blocked, suggesting that it is not a thermal effect due to the input power.  We speculate that this effect may arise from density or index changes in the hydrogen-loaded fused silica due to the UV curing process \cite{Shelby1979, Smith2001, Violakis2012}.

\section{Conclusion}

We have demonstrated single-mode photonic-crystal fibers that do not exhibit UV solarization, even at output powers of 10 mW at 280 nm and 125 mW at 313 nm.  The combination of hydrogen loading and curing with UV light injected into the fiber appears to confer long-term resistance to UV-induced color center formation.  This protocol works for fibers made from both high-OH and low-OH fused silica.  The fibers have low attenuation and bending loss.  The output mode is close to Gaussian, such that the fibers are useful for spatial filtering as well as transmission.  Additionally, the fibers can be used for single-mode propagation of disparate wavelengths from the UV to the IR.  For ion trapping applications, these fibers provide reduced beam-pointing fluctuations and stray light, as well as the ability to transfer UV light between separate optical tables.  The fiber power handling exceeds the theoretically estimated power required to perform a laser-driven fault-tolerant two-qubit gate for trapped ions (36 mW per beam at 313 nm in $^9\mathrm{Be}^+$ \cite{Ozeri2007}, by one calculation), even if two or three beams are co-propagating in a single fiber.

We note that related work on hollow-core Kagome lattice single-mode PCF for UV applications at 280 nm has been recently reported \cite{Gebert2014}.  

\section*{Acknowledgments}
Y.C. initiated hydrogen loading and curing experiments, developed the collapsing and connectorizing methods, and performed initial experiments on cured fibers.  D.H.S. and A.C.W. developed these methods further and produced and analyzed the data in this manuscript.  D.L. and D.J.W. supervised all work.  D.H.S. wrote the manuscript with input from all authors.
We thank John Gaebler, Robert J\"ordens, and Alexander Tomashuk for helpful discussions, Brian Sawyer and David Hume for manuscript comments, Till Rosenband, Sam Brewer, Jwo-Sy Chen, Kyle McKay, and David Pappas for the use of their 280 nm laser systems, Brad Sohnlein at NKT Photonics for technical advice, and the NIST Hydrogen Fuel Materials Test Facility for performing hydrogen loading of fibers.  D.H.S. acknowledges support from a National Research Council fellowship.  This paper is a contribution of NIST and is not subject to US copyright.    

\begin{thebibliography}{10}
\bibitem{footnote1}
Commercial equipment, instruments, and materials are identified in this paper in order to specify the experimental procedure accurately. Such identification does not imply recommendation or endorsement by the National Institute of Standards and Technology, nor is it intended to imply that the materials or equipment identified are necessarily the best available for the purpose.
\bibitem{footnote2}
This connectorization method is suitable for our tests, but is not sufficiently robust for fibers intended for general laboratory use.  To make patch cables for long-term experimental use, the fiber is glued into the connector in multiple locations and encased in protective furcation tubing between connectors.  The connectorized fiber tip can then be polished at a standard $8^\circ$ angle in lieu of angle cleaving.  Care must be taken when gluing the fiber into the ferrule that the collapsed region does not extend too far past the ferrule tip; otherwise, the polishing process will remove the entire collapsed region and fill the holes of the photonic crystal with polish grit and debris.  Low-stress adhesives are used to maintain good output mode quality and transmission.
\end{thebibliography}
\end{document}